\documentclass[manuscript]{aastex}

\shorttitle{Evolution of binary systems with radio pulsars}

\shortauthors{O. G. Benvenuto, M. A. De Vito \& J. E. Horvath} 

\begin{document}

\title{UNDERSTANDING THE EVOLUTION OF CLOSE BINARY SYSTEMS WITH RADIO PULSARS} 

\author{O. G. Benvenuto, M. A. De Vito}
\affil{Facultad de Ciencias Astron\'omicas y Geof\'\i sicas, Universidad 
Nacional de La Plata\\
and Instituto de Astrof\'\i sica de La Plata (IALP), CCT-CONICET-UNLP. Paseo del 
Bosque S/N (B1900FWA), 
La Plata, Argentina}

\email{obenvenu,adevito@fcaglp.unlp.edu.ar}

\and

\author{J. E. Horvath}
\affil{Instituto de Astronomia, Geof\'\i sica e Ci\^encias Atmosf\'ericas, 
Universidade de S\~ao Paulo\\
R. do Mat\~ao 1226 (05508-090), Cidade Universit\'aria, S\~ao Paulo SP, Brazil}

\email{foton@astro.iag.usp.br}

\begin{abstract} 

We calculate the evolution of close binary systems
(CBSs) formed by a neutron star (behaving as a  radio pulsar) and a normal donor
star, evolving either to helium white dwarf (HeWD) or ultra short
orbital period systems. We consider X-ray irradiation feedback
and evaporation due to  radio pulsar irradiation. 
We show that irradiation feedback leads to cyclic mass
transfer episodes, allowing CBSs to be observed in-between as binary radio pulsars under
conditions  in which standard, non-irradiated models predict the occurrence of a low mass X-ray binary. This
behavior accounts for the existence of a family of eclipsing binary
systems known as redbacks. We predict that redback 
companions should almost fill their Roche lobe, as observed in
PSR~J1723-2837. This state is also possible for systems evolving with larger orbital
periods.  Therefore, binary radio pulsars with companion star masses usually interpreted as larger than expected to
produce HeWDs may also result in such {\it quasi~-~Roche Lobe Overflow} states,
rather than hosting a carbon-oxygen WD.

We found that CBSs with initial orbital periods
$\mathrm{P_{i}<1}$~day evolve into redbacks.
Some of them produce low mass HeWDs, and a subgroup with
shorter $\mathrm{P_{i}}$ become black widows (BWs). Thus,
BWs descent from redbacks, although not all redbacks
evolve into BWs.

There is mounting observational evidence favoring
that BW pulsars are very massive ($\mathrm{\gtrsim 2\;
M_{\odot}}$). As they should be redback descendants,
redback pulsars should also be very massive, since most of the mass
is transferred before this stage.

\end{abstract}

\keywords{binaries: close --- stars: evolution ---  pulsars: general --- stars: 
neutron} 

\section{INTRODUCTION}

Close binary systems (CBSs) formed by a
neutron  star (NS)  orbiting together with a normal  donor star
experience a mass transfer episode when the donor star undergoes Roche
Lobe OverFlow (RLOF). Since then on, stars are forced to exchange  mass.
A fraction $\mathrm{\beta \leq 1}$ of the transferred  matter, carrying a
substantial amount of angular momentum, is accreted by the NS, which
becomes recycled and may reach  spin periods in the millisecond range. In
the standard treatment the initial  RLOF is a long standing episode; 
the donor star transfers a  large fraction of its mass and
the orbit is sensibly affected. During RLOF episodes, a CBS
can be detected as a low mass X-ray binary (LMXB). For CBSs with initial
orbital periods $\mathrm{P_{i} >  1}$~day, after the end of the RLOF, the
pair may be observed as a radio pulsar with a binary companion, and almost no
further  orbital evolution is expected.  Within this general picture,  if we
restrict ourselves to treat the case of stars that do not ignite helium,
binary evolution predicts a tight donor mass-orbital period\footnote{We
denote the corresponding quantities of the NS and donor star with
subscripts $\mathrm{1}$ and $\mathrm{2}$, respectively.} 
$\mathrm{P(M_{2})}$ relation (\citealt{1999A&A...350..928T};
\citealt{2002ApJ...565.1107P};  \citealt{2012MNRAS.421.2206D}). Thus, for
CBSs with $\mathrm{P_{i} >  1}$~day, standard calculations
that  lead to the formation of HeWDs predicts the
occurrence of companions to radio pulsars only in the neighborhood of $\mathrm{P(M_{2})}$
relation. Currently, companions to radio pulsars more
massive than predicted by  the $\mathrm{P(M_{2})}$ relation are
considered as CO or HeCO~WDs (\citealt{2000ApJ...530L..93T}; \citealt{2002ApJ...565.1107P}).

Two separate classes of interacting binary systems with a radio pulsar have 
been identified, they are called redbacks
and BWs. Redbacks are eclipsing binary systems with
0.1~day~$ < P <$ 1~day  and 
$\mathrm{0.2\;M_{\odot} < M_{2} < 0.4\;M_{\odot}}$. BWs  have
$P$ in the same range of values, but much
lower companion masses ($\mathrm{M_{2} < 0.05\;M_{\odot}}$).

Standard CBS models do not take into account 
neither evaporation of the donor star by radio pulsar  irradiation
\citep{1992MNRAS.254P..19S}, nor X-ray  irradiation feedback 
\citep{2004A&A...423..281B}. During a RLOF, matter
falling onto the NS produce X-ray radiation that illuminates back the
donor star, giving rise to the irradiation feedback
phenomenon. If the irradiated star has an outer  convective zone, its
structure is sensibly affected.  \citet{1985A&A...147..281V} studied
irradiated grey atmospheres  finding  that the entropy at
deep convective layers must be the same for the  irradiated and
non-irradiated portions of the star. To fulfill this condition,  the
irradiated surface is partially inhibited to release energy
emerging from its deep interior; i.e., the effective 
surface becomes smaller than $\mathrm{4 \pi  R_{2}^{2}}$  
($\mathrm{R_{2}}$ is the radius of the donor star). Irradiation
makes the evolution to depart from predicted by
standard theory. After the
onset of the RLOF, the donor star relaxes to the established conditions 
on a thermal  (Kelvin-Helmholtz) time scale,
$\mathrm{\tau_{KH}= G M^{2}_{2}/(R_{2}  L_{2}})$  ($\mathrm{G}$ is
the gravitational constant, and $\mathrm{L_{2}}$ is the luminosity of
the  donor star). In some cases, the
structure is unable to sustain the RLOF, becoming detached. Subsequent
nuclear evolution may lead the donor star to RLOF again,
undergoing a quasi cyclic behavior \citep{2004A&A...423..281B}.  
Thus, irradiation feedback may lead to the occurrence of a large number of
short lived RLOFs instead of a long standing one.  
In between, the system may reveal itself as a radio pulsar with a binary companion.
Notably, the evolution of several quantities  is only
mildly dependent on the irradiation feedback (e.g., the orbital period).

On the other hand, evaporation of the donor star, due to radio pulsar irradiation affects
the advanced evolution of CBSs that eventually reach very short orbital periods.
This effect makes  the donor star to reach very low masses and
makes these systems to
achieve orbital periods  large enough to be compatible with observations
(\citealt{2012ApJ...753L..33B};  \citealt{2013MNRAS.433L..11B}).

It is the aim of this work to explore the consequences of  
evaporation and X-ray  irradiation feedback in the evolution of these
CBSs. As stated above, CBSs  undergo cyclic RLOFs. 
This allow us to interpret that  companions to radio pulsars more
massive than predicted by the $\mathrm{P(M_{2})}$ relation may  also be
progenitors of HeWDs or ultra short orbital period systems, observationally caught 
between two consecutive RLOFs. This scenario provides a natural
explanation for the occurrence of redback  pulsars, as well as their
evolutionary link with BWs.

The reminder of the paper is as follows. In Section~\ref{sec:numeri}  we
describe our numerical evolutionary code. In
Section~\ref{sec:resultados}  we present our models, while in 
Section~\ref{sec:disc_concl} we interpret and discuss these results and
give  some concluding remarks.

\section{NUMERICAL TREATMENT} \label{sec:numeri}

We compute CBS evolution employing our code
\citep{2003MNRAS.342...50B}  updated to include irradiation feedback and
evaporation. During a RLOF episode the code solves the  structure of the
donor star simultaneously with the instantaneous mass transfer rate
$\mathrm{\dot{M}}$, the evolution of both masses and the orbital semi
axis in a {\it fully implicit} way. This strategy has the very desirable
property of numerical  stability \citep{2006A&A...445..647B} that allows
for the computation of mass transfer cycles
(\citealt{2004A&A...423..281B};  \citealt{2012ApJ...753L..33B}) presented
in this Letter. When the pair becomes  detached, the code works following the
standard Henyey technique.

Let us remark some key assumptions 
in our  calculations. We assumed that the NS  accretes a fraction $\mathrm{\beta}$
of the transferred matter up to the Eddington critical rate
$\mathrm{\dot{M}_{Edd}\approx 2 \times 10^{-8} M_{\odot}/y}$, while very
low  accretion rates ($\mathrm{\lesssim 1.3 \times 10^{-11}
M_{\odot}/y}$) are  prevented by the propeller mechanism
\citep{2008AIPC.1068...87R}. Angular momentum sinks due to
gravitational   radiation, magnetic braking and mass loss from the system
have been treated as  in \citet{2003MNRAS.342...50B}.

Irradiation feedback has been included following
\citet{1997A&AS..123..273H}.  The usual relation between luminosity, 
radius and effective temperature $\mathrm{L= 4 \pi R_{2}^{2}\; \sigma\;
T_{eff}^{4}}$ is replaced, for the case of an  irradiated photosphere by

\begin{equation}\mathrm{
L= R_{2}^{2}\; \sigma\; T_{eff,0}^{4}\; \int_{0}^{2\pi} \int_{0}^{\pi} 
G(x(\theta,\phi))\; \sin{\theta} \; d\theta\; d\phi, \label{eq:irradiando}
}\end{equation} 

\noindent where $\mathrm{\sigma}$ is the Stefan-Boltzmann constant, $\mathrm{T_{eff,0}}$ is the
effective  temperature of the non-illuminated part of the star, $\mathrm{G(x)=
\big(  T_{eff}(x)/T_{eff,0} \big)^{4} - x}$ and $\mathrm{x= F_{irr}/(\sigma\;
T_{eff,0}^{4}})$,  where $\mathrm{F_{irr}}$ is the incident irradiating flux. 
We find $\mathrm{T_{eff}(x)}$ by assuming that at
the deep layers,  the perturbations due to irradiation should vanish \citep{1985A&A...147..281V}. 
We divided the irradiated part of the stellar surface in several sectors and solved for 
$\mathrm{T_{eff}(x)}$ on each of them, and integrated over the
illuminated zone. We have ignored any heat transfer between irradiated and
non~-~irradiated zones.  This imposes a limitation to the strength of the
irradiation that our code can handle: $\mathrm{x \lesssim 10^{4}}$ \citep{2004A&A...423..281B}.
All our models are well below this limit.

We assume that the NS acts as {\it point source}, releasing an accretion
luminosity  $\mathrm{L_{acc}= G M_{1} \dot{M}_{1}/R_{1}}$ ($\mathrm{M_{1}}$, $\mathrm{R_{1}}$, and
$\mathrm{\dot{M}_{1}}$  are the mass, radius and accretion rate of the NS, 
respectively). Considering isotropy, the energy flux incident onto the
donor star is $\mathrm{F_{irr}= \alpha_{irr} L_{acc}/ 4\pi  a^{2}}$, where $\mathrm{\alpha_{irr} 
\leq 1}$ is the fraction of the incident flux that  effectively irradiates the
donor star and $\mathrm{a}$ is the orbital radius. 

Evaporation of the donor star due to radio pulsar  irradiation is included by
assuming a mass loss rate of \citep{1992MNRAS.254P..19S}

\begin{equation}\mathrm{
{\dot M}_{2,evap} = - \frac{\alpha_{evap}}{2 v_{2,esc}^{2}} L_{\rm PSR} {\biggl(
{R_{2} \over{a}} \biggr)}^{2},
\label{eq:evapora} }\end{equation}

where $\mathrm{L_{\rm PSR}}$ is the radio pulsar's spindown luminosity, $\mathrm{v_{2,esc}}$ is the escape velocity at the
donor star  surface and $\mathrm{\alpha_{evap} \leq 1}$ is an efficiency
factor. $\mathrm{L_{\rm PSR}}$ is a function of the
spin and moment of inertia evolution of the NS, that depend upon the free parameter
$\mathrm{\beta}$. These quantities,
as well as  $\mathrm{\alpha_{evap}}$, are uncertain. So, as a  first approximation we consider the product
$\mathrm{\alpha_{evap} L_{\rm PSR}}$ as time  independent adopting a
range of plausible values. 

Previous calculations \citep{2012ApJ...753L..33B} indicate that 
evaporation is relevant for CBSs with very short orbital periods only,
especially  after these CBSs reach their minimum value. Because of this
reason, we considered evaporation since the minimum period is reached onwards.

``Radio ejection'' may be relevant for these CBSs. It has  been
proposed (\citealt{1989ApJ...336..507R}; \citealt{2001ApJ...560L..71B}) 
that the pressure due to radio pulsar
irradiation may  inhibit material lost by the donor star to be
accreted by the NS. If ``radio  ejection'' operates, it precludes
irradiation feedback to occur. 
In addition, we have neglected the presence of an accretion
disk surrounding the NS. 
\citet{2012A&A...537A.104V} studied the evolution of ultracompact
X-ray binary systems, considering an
accretion disk, allowing for the occurrence of short time
scales disk instabilities, modulating the long term evolution computed here.
Including ``radio ejection'' and/or accretion disks is beyond the
scope of this Letter.

\section{RESULTS} \label{sec:resultados}

\begin{figure} \begin{center}
\includegraphics[scale=.50,angle=270]{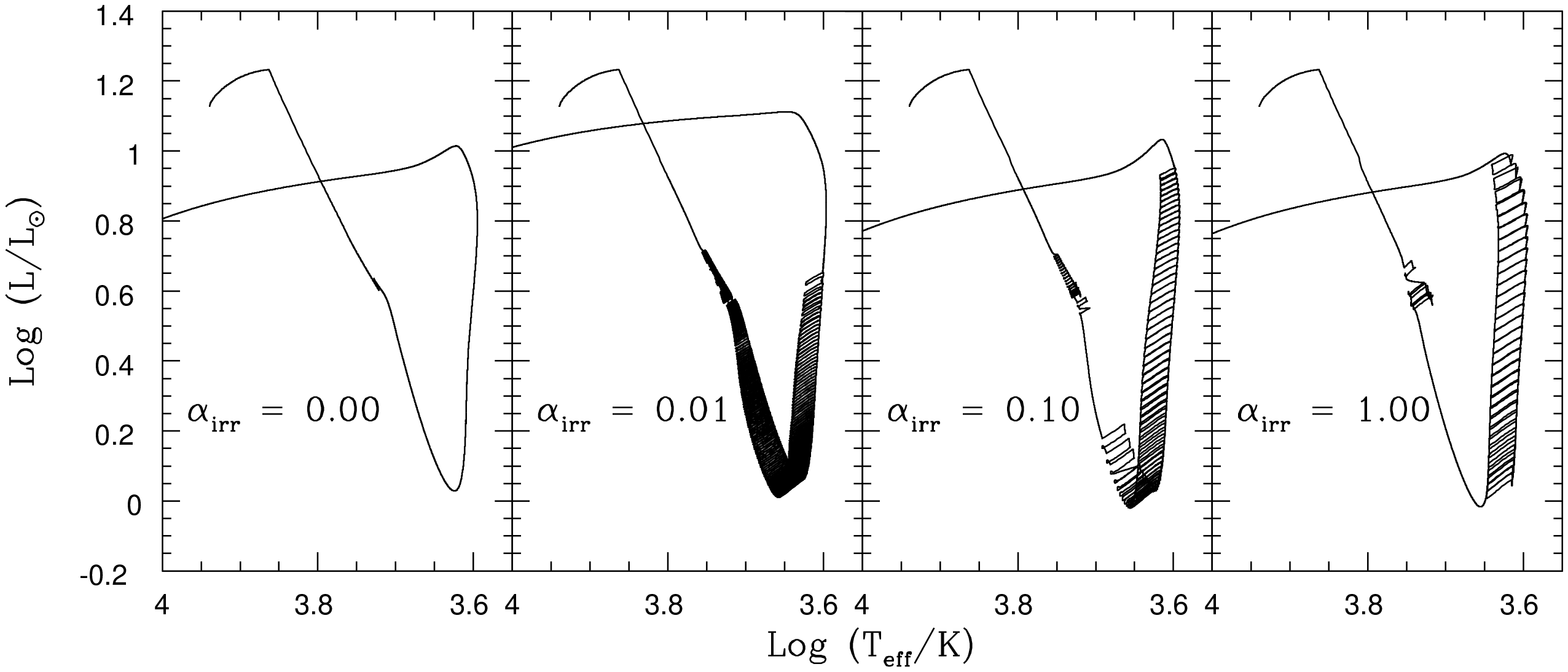} \caption{Early
evolution of CBSs that initially have a normal, solar composition
$\mathrm{M_{2}= 2\; M_{\odot}}$ star and a ``canonical'' NS
($\mathrm{M_{1}= 1.4\; M_{\odot}}$)  on a  circular 1~day orbit.
The effective temperature of the horizontal axis  corresponds to
the non~-~irradiated portion of the surface. Left panel  shows
the evolution of the system without irradiation feedback,
while the others represent the cases of  different strengths of
irradiation, denoted by the value of $\mathrm{\alpha_{irr}}$. The non~-~irradiated model suffers from one long
standing RLOF. In contrast, irradiated models
undergo a large number of  RLOFs, separated by detached
stages (See  Fig.~\ref{fig:pulsos_mdot}).  Evaporation is not important
at the evolutionary stages depicted here. \label{fig:tracks_irrad}}
\end{center}  \end{figure}

\begin{figure} \begin{center}
\includegraphics[scale=.50,angle=0]{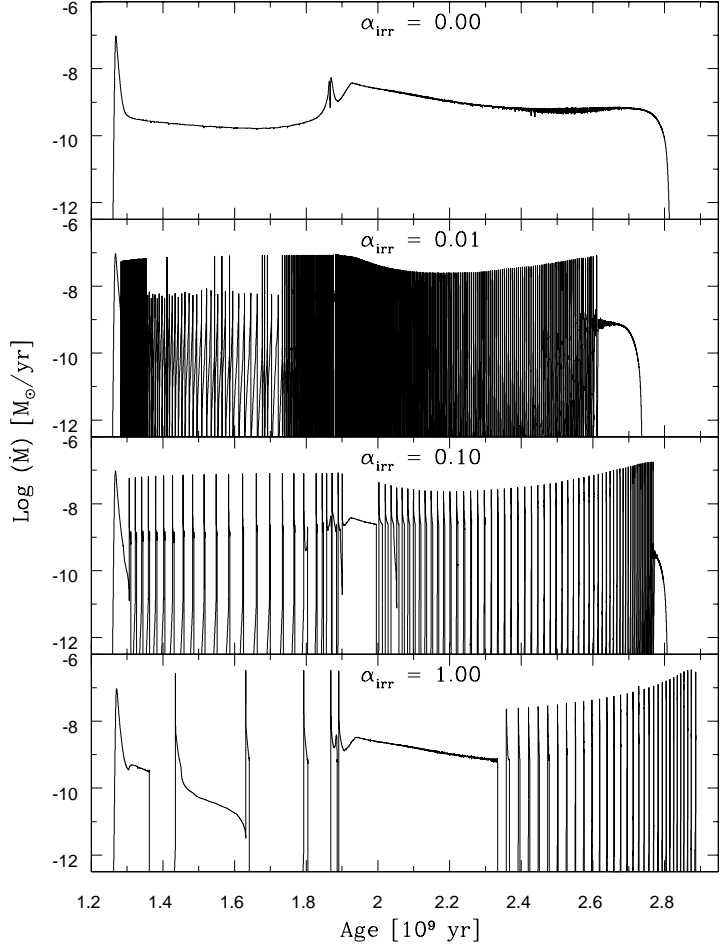} \caption{Evolution of
the mass  transfer rate for the CBSs included in
Fig.~\ref{fig:tracks_irrad}. Upper  panel shows the case of the
non~-~irradiated model, while the others correspond to  the cases of
different strengths of irradiation, denoted by the value of
$\mathrm{\alpha_{irr}}$. When the donor star exhausts its
hydrogen core (at an  age of $\approx$~1.9~Gyr), mass transfer becomes
momentarily stable. The NS accretion rate is limited by the
Eddington rate ($\mathrm{\approx 2 \times 10^{-8} 
M_{\odot}/y}$). \label{fig:pulsos_mdot}}
\end{center}  \end{figure}

We computed CBSs that initially have  donor masses from
1.00~$\mathrm{M_{\odot}}$ to 3.50~$\mathrm{M_{\odot}}$ with steps of 
0.25~$\mathrm{M_{\odot}}$; a ``canonical'' NS ($\mathrm{M_{1}=1.4\; M_{\odot}}$); and  $P_{i}$ of  0.50, 0.75, 1.00, 
1.50, 3.00, 6.00, and 12~days. We set $\mathrm{\beta=0.5}$
for all calculations. At  least for models without
irradiation, donor  evolution is almost
independent of $\mathrm{\beta}$. If $\beta<0.5$, 
irradiation will be weaker and NSs lighter. 
Pulsed mass transfer is found for 
$\mathrm{\alpha_{irr}>0}$. Thus,  lowering $\mathrm{\beta}$ should not change
the behavior of the models qualitatively.  We consider $\mathrm{\alpha_{irr}}=$
0.00, 0.01, 0.10, and  1.00; and
$\mathrm{\alpha_{evap}\; L_{\rm PSR}= 0.00, 0.04, 0.20}$, 
and $\mathrm{1.00\; L_{\odot}}$. Typical early evolutionary tracks
are  presented in Fig.~\ref{fig:tracks_irrad}, while the
evolution of $\mathrm{\dot{M}}$ for these CBSs is
shown in Fig.~\ref{fig:pulsos_mdot}. 

As stated above, during mass transfer, CBSs switch from accretion (LMXBs) to 
detachment conditions. The number of RLOFs is  strongly dependent on
$\mathrm{\alpha_{irr}}$. The non~-~irradiated model (upper
panel of  Fig.~\ref{fig:pulsos_mdot}) behaves is an LMXB in almost
the entire time period of  Fig.~\ref{fig:pulsos_mdot}.  In contrast,
irradiated  models allow for its detection either as LMXB or radio pulsar with a companion (see
the other panels of  Fig.~\ref{fig:pulsos_mdot}) in  detached state. While the number of RLOFs
is strongly dependent on  $\mathrm{\alpha_{irr}}$, the duty
cycle for detecting a LMXB or a radio pulsar is nearly constant.   In the
cases of $\mathrm{\alpha_{irr}= 0.10}$ and 1.00, the cycles of mass
transfer  and detachments are visible in
Fig.~\ref{fig:pulsos_mdot}. For $\mathrm{\alpha_{irr}=
0.01}$  the average width of the cycle is $\mathrm{\approx 1}$~Myr on
RLOF and $\approx 2$~Myr detached.

In Fig.~\ref{fig:plano_masper} we present the evolution of CBSs
together with available observed systems in the  plane
$\mathrm{M_{2}}$ vs. $\mathrm{P}$.  We restrict the data sample to those
systems that contain one NS which have also undergone at least one RLOF. Those CBSs should
have suffered strong tidal dissipation, and their orbits should be almost
circular \citep{1977A&A....57..383Z}. The radio pulsar data has been
taken  from the ATNF data
base\footnote{\url{www.atnf.csiro.au/research/pulsar/psrcat}} 
\citep{2005AJ....129.1993M},  Freire's
page\footnote{\url{www.naic.edu/~pfreire/GCpsr.html}} (altough they 
correspond to other chemical abundances) and Fermi gamma 
ray sources (\citealt{2011ApJ...732...47C};
\citealt{2011AIPC.1357...40H};  \citealt{2011ApJ...727L..16R};
\citealt{2012ApJ...748L...2K}; \citealt{2013ApJ...763L..13R}; 
\citealt{2013ApJ...769..108B}). If for a given object more than one data
sample is available, we have plotted the larger value reported for the mass. 

In Fig.~\ref{fig:plano_masper}, CBSs with $\mathrm{P_{i} \gtrsim 1}$~day
evolve towards the  $\mathrm{P(M_{2})}$ relation. 
If irradiation feedback is considered, there are stages at
which the CBS is detached while the track passes trough a
region populated by several systems.  CBSs
that do {\it not} reach BW region form a
low mass, HeWD. We found that during the whole unstable mass transfer cycle the 
donor stars almost fill their Roche lobe.
This is the result of the interplay between irradiation and orbital evolution.
We call this donor stars state as quasi-RLOF hereafter.
This represents  an alternative evolutionary
status for these companion objects. A very recently detected  redback system, PSR
J1723-2837 \citep{2013ApJ...776...20C} support these results. Observations
indicate that the Roche lobe of the donor star is  almost
completely filled, as predicted by these calculations.  To be sure, there are other
paths to arrive to the right of the  $\mathrm{P(M_{2})}$ relation. These may actually be
CO, or HeCO WDs  (see
\citealt{2000ApJ...530L..93T}; \citealt{2002ApJ...565.1107P}). Our results indicate that another 
quite different evolutionary path exists for these objects to occur.

In Fig.~\ref{fig:zona_redbacks} we show the redbacks region of Fig.~\ref{fig:plano_masper}, where 
we include all computed models together with the
observed systems. Redback systems  with $P\gtrsim 6$~h should
form a low mass HeWD-radio pulsar system. 
For shorter  $P$, they are clearly BW progenitors.
Many evolutionary  tracks converge, evolving to the BW region marked in  Fig.~\ref{fig:plano_masper}. In short,
to become a BW it is necessary to have had 
characteristics similar to those of redback previously. Nevertheless, not
all redbacks evolve to BW.  Notice that, in the redbacks region
of Fig.~\ref{fig:zona_redbacks}, irradiated  models undergo cyclic mass
transfer; so, we expect the detection of  some binary systems
as containing a radio pulsar. We find
pulsed mass transfer at redbacks conditions for
all considered initial donor masses and short $P$ values.  Again,  pulsed mass transfer is found especially for
$\mathrm{\alpha_{irr}= 0.01}$, although it is also present
for stronger irradiation  regimes.

Another relevant item regarding redbacks and BW, is the mass of the underlying NS. A necessary
condition for a CBS to  become a redback is its short $P_{i}$. 
For  such CBSs, mass transfer occurs at low
rates. Thus, most of the matter is accreted by the  NS resulting 
significantly more massive. While the mass growth of the  NSs depends on
$\mathrm{\beta}$, within the adopted assumptions we 
found redbacks with masses of 2.4~$\mathrm{M_{\odot}}$. This is
consistent with  recent reports (\citealt{2011ApJ...728...95V};
\citealt{2012ApJ...760L..36R}) claiming high NS masses for some BW.

The BWs region is reached by systems that evolved 
trough the redback region (see Fig.~\ref{fig:plano_masper}).  
Evolutionary tracks with and without irradiation feedback are hardly distinguishable.
However, this is not the case
regarding  evaporation. Models without evaporation evolve at very low
$P$,  in conflict with observations. Evaporation makes the  trajectories to bend up (in
Fig.~\ref{fig:plano_masper}) reaching the required range of
$\mathrm{P}$  values at ages shorter than the age of the Universe. This is independent from the 
earlier operation of irradiation feedback, and should even  occur if
``radio ejection'' prevents NS accretion. 
BW companions are semi~-~degenerate;  thus, evaporation makes
the star to swell  and the orbit to widen fast enough to detach  from its
Roche Lobe  (\citealt{2012ApJ...753L..33B};
\citealt{2013MNRAS.433L..11B}). The
larger the $L_{\rm PSR}$,  the steeper the trajectory. Interestingly,  BW 
companions  become fully convective for $\mathrm{Log(T_{eff}/K)
\lesssim 3.4}$ independently  of the evaporation rate.

\section{DISCUSSION AND CONCLUSIONS} \label{sec:disc_concl}

We computed the evolution of  CBSs formed by a NS (behaving as a radio pulsar) together with
a normal donor star, for a range of donor masses for which we  expect as
a final product a low mass HeWD or an ultracompact system. We 
included X-ray irradiation feedback and evaporation due to radio pulsar
irradiation.  These important ingredients allow for the occurrence of
radio pulsars with companions in evolutionary stages at which donor stars are
undergoing episodic mass transfers and almost filling their Roche lobe (the quasi-RLOF state),
being widely different from a WD  star. This seems to fit the case of
J1723-2837 \citep{2013ApJ...776...20C} which remains otherwise hard to interpret.

We assumed that during RLOFs, the NS
acts as  an X-ray point  source 
\citep{1997A&AS..123..273H}. This represents the simplest configuration,
but not the only possible one \citep{2008NewAR..51..869R}. For example, the
X-ray flux may come from an accretion disk. In any case, a sequence of RLOF and detachment cycles are
expectable. Notice that for all the values of
$\mathrm{\alpha_{irr}>0}$ considered here we find a  cyclic behavior for
the mass transfer. Apparently, donor stars with convective envelopes are very  sensitive to
irradiation feedback, making the scenario presented  here a robust one.
This is so, provided ``radio ejection'' is
not operative; otherwise it would inhibit accretion onto
the NS and irradiation feedback too, but evaporation should occur.

Part of our models evolve to redback conditions.
This is so for the whole range of initial masses and short
initial orbital periods  ($\mathrm{P<1}$~day). For all the values of
$\mathrm{\alpha_{irr}>0}$ we find cyclic mass transfer  episodes. A
fraction of them form a low mass HeWD, but  others
become BWs progenitors. Therefore, our calculations indicate that all
BWs should have passed through the redbacks region (although
at such stage they are not necessarily classified as redbacks if eclipses are not seen).
\citet{2013ApJ...775...27C} studied this problem, based on models including evaporation.
They found that only fine tuning the evaporation rate for systems with $P < 4$~h, 
it is possible to find BWs as redback descendants. This result is very different 
from the scenario described by our models.

As stated in Section~\ref{sec:resultados}, some BWs have been reported to 
harbour massive NSs. In our scenario in which BWs  are redback descendants, 
we infer that the NSs that behave as redbacks should be massive too. This is 
because most of the accretion onto the NSs should have occurred previous to 
reaching redback conditions. We  found CBSs 
corresponding to redbacks with NS masses as high as 2.4~$\mathrm{M_{\odot}}$.

In summary, we have calculated a set of evolutionary tracks showing the relevance
of irradiation/evaporation stages for interacting CBSs. Although other relevant
ingredients may still be lacking, we believe there is enough evidence that a unified
picture of the redback/BW evolution is closely related to the consideration
of these physical effects, allowing an interpretation of present observational
evidence.

\begin{figure}  \begin{center} 
\includegraphics[scale=.40,angle=0]{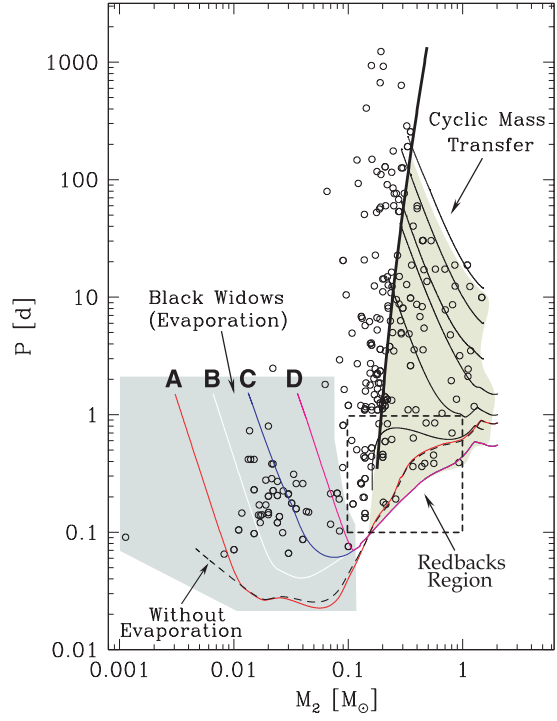} \caption{Donor mass
vs.  orbital period for CBSs, together with the corresponding 
observational data. Open dots denote {\it minimum} masses for radio pulsar 
companions.  The $\mathrm{P(M_{2})}$ relationship is denoted with a thick
solid line.  On the right, thin solid lines denote the evlution of CBSs
with initial donor masses and orbital periods of
1.50~$\mathrm{M_{\odot}}$ with $\mathrm{P_{i}= 0.75,  3.0, 6.0}$, and
$\mathrm{12.0}$~days; and 2.00~$\mathrm{M_{\odot}}$ with $\mathrm{P_{i}=
1.00}$, and  $\mathrm{1.50}$~days. All of them end close to the
$\mathrm{P(M_{2})}$ curve. Tracks  reaching the BWs region
(shaded on the left) correspond to an initial donor mass of 
2.00~$\mathrm{M_{\odot}}$. Three of them have $\mathrm{P_{i}= 0.55}$~days
and $\mathrm{\alpha_{evap}\;  L_{\rm PSR}= 0.04, 0.20}$, and
$\mathrm{1.00\; L_{\odot}}$ (B, C, and D respectively) and 
another with $\mathrm{P_{i}= 0.85}$~days and $\mathrm{\alpha_{evap}\;
L_{\rm PSR}=  0.04\;L_{\odot}}$ (A). The shaded region on the
right denotes conditions on which pulsed mass transfer occurs. The
BWs region is reached by systems that previously evolved trough
the  redback region. Models without
evaporation  (see the long dashed line representing the case of
2.0~$\mathrm{M_{\odot}}$ with $\mathrm{P_{i}=  0.85}$~days) evolve at too low
orbital periods, in conflict with observations. 
\label{fig:plano_masper}}  \end{center} \end{figure}
\begin{figure}  \begin{center} 
\includegraphics[scale=.50,angle=0]{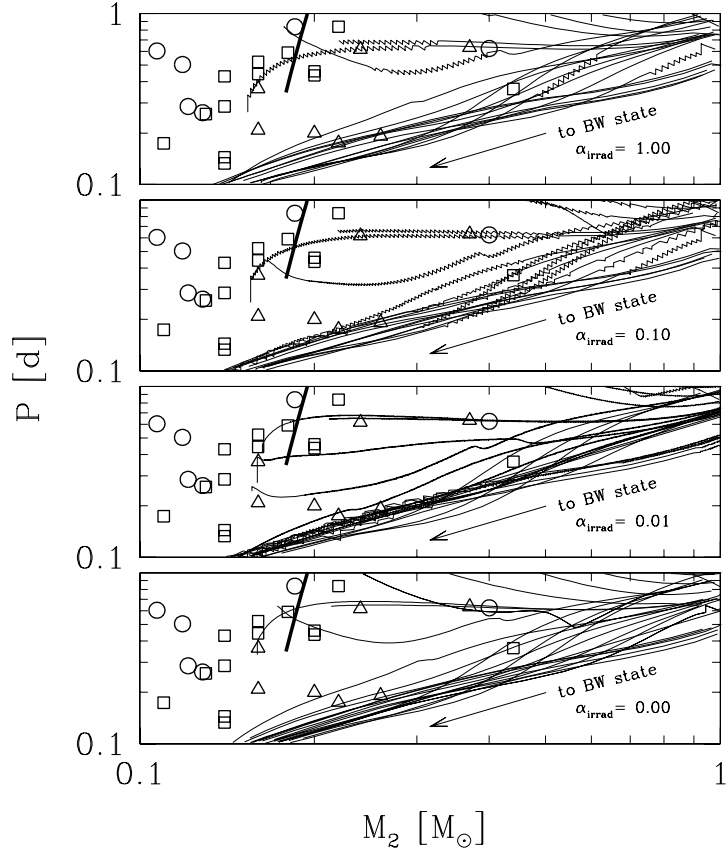} \caption{The
region of Fig.~\ref{fig:plano_masper} where redbacks are
located. Each panel  corresponds to different irradiation conditions.  
Hollow circles,  triangles and squares
correspond to observed systems taken for  the ATNF data base
\citep{2005AJ....129.1993M}, \citet{2013IAUS..291..127R}, and  Freire's
home page, respectively. The thick solid line represents the low mass 
end of the mass-period relation.
\label{fig:zona_redbacks}}  \end{center} \end{figure}

\acknowledgments We deeply acknowledge our anonymous referee for a very 
constructive report that allowed us to markedly improve the original version
of  this paper. O.G.B. thanks Leonardo Benvenuto for his assistance in
the preparation of  Fig.~\ref{fig:plano_masper}. O.G.B. is member of the
Carrera de Investigador of the CIC-PBA  Agency and M.A.D.V. is member of
the Carrera del Investigador Cient\'{\i}fico,  CONICET, Argentina. J.E.H.
has been supported by Fapesp (S\~ao Paulo, Brazil) and CNPq,  Brazil
funding agencies.


\clearpage

%
%

\clearpage

\end{document}